\documentclass[prc,twocolumn,showpacs,preprintnumbers,amsmath,amssymb,superscriptaddress,floatfix,nofootinbib]{revtex4}
\usepackage{graphicx}
\usepackage{amsmath}
\usepackage{amsfonts}
\usepackage{amssymb}%

\newcommand{\Slash}[1]{\ooalign{\hfil/\hfil\crcr$#1$}}

\begin{document}

\title{Role of the $\Delta^*(1940)$ in the $\pi^+ p \to K^+ \Sigma^+(1385)$ and $pp \to n K^+ \Sigma^+(1385)$ reactions }
\author{Ju-Jun Xie} \email{xiejujun@impcas.ac.cn}
\affiliation{Institute of Modern Physics, Chinese Academy of
Sciences, Lanzhou 730000, China} \affiliation{Research Center for
Hadron and CSR Physics, Institute of Modern Physics of CAS and
Lanzhou University, Lanzhou 730000, China} \affiliation{State Key
Laboratory of Theoretical Physics, Institute of Theoretical Physics,
Chinese Academy of Sciences, Beijing 100190, China}
\author{En Wang} \email{En.Wang@ific.uv.es}
\affiliation{Departamento de F\'\i sica Te\'orica and IFIC, Centro
Mixto Universidad de Valencia-CSIC, Institutos de Investigaci\'on de
Paterna, Aptd. 22085, E-46071 Valencia, Spain}
\author{Bing-Song Zou} \email{zoubs@itp.ac.cn}
\affiliation{State Key Laboratory of Theoretical Physics, Institute
of Theoretical Physics, Chinese Academy of Sciences, Beijing 100190,
China}

\begin{abstract}

The $p p \to n K^+ \Sigma^+(1385)$ reaction is a very good isospin
$3/2$ filter for studying $\Delta^{++*}$ resonance decaying to
$K^+\Sigma^+(1385)$. Within the effective Lagrangian method, we
investigate the $\Sigma(1385)$ (spin-parity $J^P = 3/2^+$) hadronic
production in the $\pi^+ p \to K^+ \Sigma^+(1385)$ and $p p \to n
K^+ \Sigma^+(1385)$ reactions. For $\pi^+ p \to K^+ \Sigma^+(1385)$
reaction, in addition to the ``background" contributions from
$t$-channel $K^{*0}$ exchange, $u$-channel $\Lambda(1115)$ and
$\Sigma^0(1193)$ exchange, we also consider the contribution from
the $s$-channel $\Delta^*(1940)$ resonance, which has significant
coupling to $K\Sigma(1385)$ channel. We show that the inclusion of
the $\Delta^*(1940)$ resonance leads to a fairly good description of
the low energy experimental total cross section data of $\pi^+ p \to
K^+ \Sigma^+(1385)$ reaction. Basing on the study of $\pi^+ p \to
K^+\Sigma^+(1385)$ reaction and with the assumption that the
excitation of $\Delta^*(1940)$ resonance dominants the $pp \to n K^+
\Sigma^+(1385)$ reaction, we calculate the total and differential
cross sections of the $p p \to n K^+ \Sigma^+(1385)$ reaction. It is
shown that the new experimental data support the important role
played by the $\Delta^*(1940)$ resonance with a mass in the region
of $1940$ MeV and a width of around $200$ MeV. We also demonstrate
that the invariant mass distribution and the Dalitz Plot provide
direct information of the $\Sigma^+(1385)$ production, which can be
tested by future experiments.

\end{abstract}

 \maketitle

\section{Introduction}

Study of the spectrum of isospin $3/2$ $\Delta^{++}(1232)$ excited
states is one of the most important issues in hadronic physics and
is attracting much attention because it is the most experimentally
accessible system composed of three identical valence quarks.
However, our knowledge on these resonances mainly comes from old
$\pi N$ experiments and is still very poor~\cite{klempt,pdg2012}. In
the energy region around or above $2.0$ GeV, there are still many
theoretical predictions of ``missing $\Delta^*$ states", within the
constituent quark~\cite{capstick2000} or chiral
unitary~\cite{Sarkar:2009kx,Oset:2009vf,Gamermann:2011mq,Sun:2011fr}
approaches, which have so far not been observed. Searching for these
``missing $\Delta^*$ states" from other production processes is
necessary~\cite{Xie:2007vs,Doring:2010ap}. A possible new excellent
source for studying these $\Delta^*$ resonance comprises the $\pi^+
p \to K^+ \Sigma^+(1385)$ and $pp \to nK^+\Sigma^+(1385)$ reactions,
which have a special advantage since there is no contributions from
isospin $1/2$ nucleon resonances because of the isospin and charge
conservations. In addition, those reactions require the creation of
an $\bar{s}s$ quark pair. Thus, a thorough and dedicated study of
the strangeness production mechanism in those reactions has the
potential to gain a deeper understanding of the interaction among
strange hadrons and also the nature of the $\Delta^*$ resonances.

In analogy to the $\Delta(1232)$ as first-excited state of the
nucleon, the $\Sigma(1385)$ is the first-excited state of the
$\Sigma(1193)$ hyperon and has a spin-parity of $3/2^+$ and isospin
1. This resonance is considered as a standard quark triplet and
cataloged in the baryon decuplet, but its vicinity to the
$\Lambda(1405)$ state in the mass spectrum correlates the study and
the understanding of the two resonances. On the other hand, a
$\Sigma$ state, $\Sigma(1380)$ (spin-parity $J^P = 1/2^-$) with mass
about $1380$ MeV, was predicted in the framework of the
diquark-diquark-antiquark
picture~\cite{Zhang:2004xt,Wu:2009tu,Wu:2009nw}. This new state will
make effects in the production of $\Sigma(1385)$ and then the
analysis of the $\Sigma(1385)$ resonance suffers from the
overlapping mass distributions and the common $\pi \Lambda(1115)$
decay mode.

There were pioneering measurements in the 1970s, the first $p p \to
n K^+ \Sigma^+(1385)$ cross sections in the high energy region, with
beam momentum p$_{\rm lab}$ = 6 GeV, were reported in
Ref.~\cite{Klein:1970ri}. Recently, this reaction was examined at
$3.5$ GeV beam energy by HADES
Collaboration~\cite{Agakishiev:2011qw}. The results of angular
distributions of the $\Sigma^+(1385)$ in different reference frame
show that there could be contribution from an intermediate
$\Delta^*$ resonance via the decay of $\Delta^{++*} \to K^+
\Sigma^+(1385)$. Thus, the study of the possible role played by
$\Delta^*$ resonances in the available new data from the HADES
Collaboration is timely and could shed light on the complicated
dynamics that governs the spectrum of these $\Delta^*$ states.

The theoretical activity has also run in parallel. Thinking of the
$p p \to n K^+ \Sigma^+(1385)$ reaction, the one-boson exchange
model can be considered. By using this frame, several theoretical
calculations by considering the $\pi$ exchange
diagrams~\cite{Klein:1970ri}, the $\pi$ and $K$ exchange
diagrams~\cite{Ferrari:1969fr}, and the intermediate $\Delta^{++}$
excitation~\cite{Chinowsky:1968rn}, exist for describing the old and
high energy data of Ref.~\cite{Klein:1970ri}. These theoretical
studies have traditionally been limited by the lack of knowledge on
the $\Delta^* \Sigma(1385)K$ coupling strength and also the new
experimental measurements from HADES~\cite{Agakishiev:2011qw}.

In this work, we study the $\pi^+ p \to K^+ \Sigma^+(1385)$ and $p p
\to n K^+ \Sigma^+(1385)$ reactions within the effective Lagrangian
method by examining the important role of the $\Delta^*$ resonances
in these reactions. For the $\pi^+ p \to K^+ \Sigma^+(1385)$
reaction, in addition to the ``background" contributions from the
$t$-channel $K^{*0}$ exchange, and $u$-channel $\Lambda(1115)$ and
$\Sigma^0(1193)$ hyperon pole terms, we also study possible
contributions from $\Delta^*$ resonances in the $s$-channel. Based
on the results obtained from $\pi^+ p \to K^+ \Sigma^+(1385)$
reaction, we tend to study the role of $\Delta^*$ resonances in the
$p p \to n K^+ \Sigma^+(1385)$ reaction with the assumption that the
production mechanism is due to the $\pi^+$-meson exchange with the
aim of describing the new experimental data reported by HADES.
Unfortunately, the information about the strong coupling of
$\Delta^*K\Sigma(1385)$ is scarce~\cite{pdg2012}. Thus, it is
necessary to rely on theoretical schemes, such that of
Refs.~\cite{Capstick:1992uc,simonprd58} based on a quark model (QM)
for baryons. Among the possible $\Delta^*$ resonances, we have
finally considered only the two-star $D$-wave $J^P=3/2^-$
$\Delta^*(1940)$, which is predicted to have visible
contribution~\cite{simonprd58} to the $K \Sigma(1385)$ production.
Indeed, in
Refs.~\cite{Oh:2007jd,Gao:2010hy,Chen:2013vxa,He:2013ksa}, the
contribution from a $\Delta^*$ resonance with spin-parity $3/2^-$
and mass around $2$ GeV was studied in the $\gamma p \to K^+
\Sigma^0(1385)$ reaction. They all found that this $\Delta^*$
resonance has a significant coupling to $K \Sigma(1385)$ channel and
plays an important role in the reaction of $\gamma p \to K^+
\Sigma^0(1385)$.\footnote{In Refs.~\cite{Gao:2010hy,Chen:2013vxa},
the role played by the pentaquark state, $\Sigma^*(1380)$
(spin-parity $J^P = 1/2^-$), is also studied. But the knowledge on
this state is very scarce. We thus leave the investigation on the
role of this new state to a future study.} Although the
$\Delta^*(1940)$ resonance is listed in the Particle Data Group
(PDG) book, the evidence of its existence is poor or only fair and
further work is required to verify its existence and to know its
properties, accordingly, its total decay width and branching ratios
are not experimentally known, either. In this respect, the HADES
measurements could be used to determine some properties of this
resonance.

To end this introduction, we would like to mention that in
Refs.~\cite{Oh:2007jd,Gao:2010hy,Chen:2013vxa,He:2013ksa}, the role
played by another $\Delta^*$ resonance, $\Delta^*(2000)$
(spin-parity $J^P = 5/2^+$), in the $\gamma p \to K^+
\Sigma^0(1385)$ reaction has been also studied. In these works, it
is shown that the $\Delta^*(2000)$ resonance has a dominant
contribution. However, it is pointed out, in Ref.~\cite{Xie:2011uw},
that the nominal mass of the $\Delta^*(2000)$ resonance does not
correspond in fact to any experimental analysis but to an estimation
based on the value of masses ($\sim 1740$ and $2200$ MeV) extracted
from different data analysis~\cite{pdg2012}. From the results
obtained in Ref.~\cite{Xie:2011uw} we may conclude that the two
distinctive resonances, $\Delta^*(\sim 1740)$ and $\Delta^*(\sim
2200)$, should be cataloged instead of $\Delta^*(2000)$. We thus
will not consider the contribution from $\Delta^*(2000)$ resonance
in the present work.

In the next section, we will show the formalism and ingredients in
our calculation, then numerical results and discussions are
presented in Sect. III. A short summary is given in the last
section.

\section{Formalism and ingredients}

The combination of effective Lagrangian approach and isobar model is
an important theoretical tool in describing the various processes in
the region of resonance produced. In this section, we introduce the
theoretical formalism and ingredients to calculate the
$\Sigma(1385)$ ($\equiv \Sigma^*$) hadronic production in $\pi^+ p
\to K^+ \Sigma^+(1385)$ and $p p \to n K^+ \Sigma^+(1385)$ reactions
within the effective Lagrangian approach and isobar model.

Because we only consider the tree diagrams for the $\pi^+ p \to K^+
\Sigma^+(1385)$ and $pp \to nK^+ \Sigma^+(1385)$ reactions, the
total scattering amplitudes have not taken into account of the
unitary requirements, which may be important for extracting the
parameters of the baryon resonances from the analysis of the
experimental data~\cite{Kamano:2009im,Suzuki:2009nj}, especially for
those reactions involving many intermediate couple channels and
three-particle final states~\cite{Kamano:2008gr,Kamano:2011ih}. On
the other hand, we know that it is difficult to really apply the
unitary constraints in the three body cases, which need to include
the complex loop diagrams~\cite{Kamano:2011ih,MartinezTorres:2009cw,
alberto}, and the extracted rough parameters for the major
resonances still provide useful information, hence we will leave it
to further studies. Nevertheless, our model used in the present work
can give a reasonable description of the experimental data in the
considered energy region. Meanwhile, our calculation offers some
important clues for the mechanisms of the $\pi^+ p \to K^+
\Sigma^+(1385)$ and $pp \to nK^+ \Sigma^+(1385)$ reactions and makes
a first effort to study the role of the $\Delta^*(1940)$ resonance
in the relevant reactions.

\subsection{Feynman diagrams and effective interaction Lagrangian densities} \label{feylag}

\begin{figure*}[htbp]
\begin{center}
\includegraphics[scale=0.5]{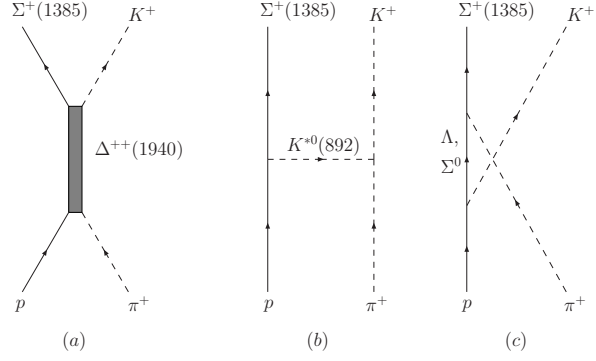}
\caption{Feynman diagrams for $\pi^+ p \to K^+ \Sigma^+(1385)$
reaction. The contributions from $s$-channel $\Delta^{++}(1940)$
resonance, $t$-channel $K^{*0}$ exchange, and $u$-channel
$\Lambda(1115)$ and $\Sigma^0(1193)$ exchange are considered.}
\label{pipdiagram}
\end{center}
\end{figure*}

\begin{figure*}[htbp]
\begin{center}
\includegraphics[scale=0.6]{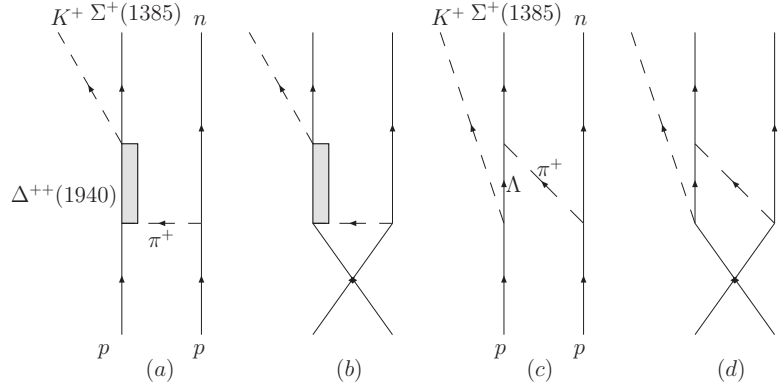}
\caption{Feynman diagrams for $p p \to n K^+ \Sigma^+(1385)$
reaction. The diagrams (a) and (c) show the direct processes, while
(b) and (d) show the exchange processes.} \label{ppdiagram}
\end{center}
\end{figure*}

The basic tree level Feynman diagrams for the $\pi^+ p \to K^+
\Sigma^+(1385)$ and $p p \to n K^+ \Sigma^+(1385)$ reactions are
depicted in Fig.~\ref{pipdiagram} and Fig.~\ref{ppdiagram},
respectively. For the $\pi^+ p \to K^+ \Sigma^+(1385)$ reaction, in
addition to the ``background" diagrams, such as $t$-channel $K^{*0}$
exchange [Fig.~\ref{pipdiagram}(b)], $u$-channel $\Lambda(1115)$ and
$\Sigma^0(1193)$ exchange [Fig.~\ref{pipdiagram}(c)], we also
consider the $s$-channel $\Delta^{++*}(1940)$ ($\equiv \Delta^*$)
resonance excitation process [Fig.~\ref{pipdiagram}(a)].

In Fig.~\ref{ppdiagram}, we show the tree-level Feynman diagrams for
$p p \to n K^+ \Sigma^+(1385)$ reaction. The diagram
Fig.~\ref{ppdiagram}(a) and Fig.~\ref{ppdiagram}(c) show the direct
processes, while Fig.~\ref{ppdiagram}(b) and Fig.~\ref{ppdiagram}(d)
show the exchange processes. It is assumed that the production of
the $K^+\Sigma^+(1385)$ passes mainly through the
$\Delta^{++*}(1940)$, which has a significant coupling to
$K\Sigma(1385)$. In this case, the $t$-channel $K^{*0}$ exchange and
$u$-channel $\Sigma^0(1193)$ exchange processes are neglected since
their contributions are small, which will be discussed below.

For the $\pi^+ p \to K^+ \Sigma^+(1385)$ reaction, to compute the
contributions of those terms shown in Fig.~\ref{pipdiagram}, we use
the interaction Lagrangian densities as in
Refs.~\cite{Oh:2007jd,Zou:2002yy,Kim:2012pz},
\begin{eqnarray}
{\mathcal L}_{\pi N \Delta^*}  &=& \frac{g_{\pi N \Delta^*}}{m_{\pi}}
\bar{\Delta}^{* \mu} \gamma_5 (\partial_{\mu} \vec\tau \cdot \vec\pi) N\,+{\rm h.c.},  \label{pinnstar} \\
\mathcal{L}_{K \Sigma^* \Delta^*} &=& \frac{g_1}{m_{K}}
\bar{\Sigma}^*_{\mu} \gamma_{\alpha} (\partial^{\alpha} K)
\Delta^{*\mu} \, + \nonumber \\
&& \frac{i g_2}{m_{K}^2} \bar{\Sigma}^*_{\mu} \left (\partial^{\mu}
\partial_{\nu} K\right)  \Delta^{*\nu} \,+{\rm h.c.}, \label{eq:eqknstar}
\end{eqnarray}
for the $s$-channel $\Delta^{*}(1940)$ processes, and
\begin{eqnarray}
{\mathcal L}_{K^* N \Sigma^*}  &=&  i \frac{g_{K^* N
\Sigma^*}}{2m_{N}} \bar{N} \gamma^{\nu} \gamma_5 \Sigma^{*\mu}
(\partial_{\mu} K^*_{\nu} - \partial_{\nu}K^*_{\mu}) \nonumber \\ && +  {\rm h.c.}, \\
\label{kstarnlamstar} \mathcal{L}_{K^* K \pi}   &=&  g_{K^*K\pi}
[\bar{K} (\partial^{\mu} \vec\tau \cdot \vec\pi) -
(\partial^{\mu}\bar{K})\vec\tau \cdot \vec\pi ]K^*_{\mu} \nonumber
\\ &&  + {\rm h.c.},  \label{eq:kstarkpi}
\end{eqnarray}
for the $t$-channel $K^{*0}$ exchange process, while
\begin{eqnarray}
{\mathcal L}_{K N \Sigma/\Lambda}  &=& -i g_{K N \Sigma/\Lambda}
\bar{N} \gamma_5 K \Sigma/\Lambda \,+ {\rm h.c.}, \\
\label{knnsigma} \mathcal{L}_{\Sigma^* \pi \Sigma/\Lambda} &=&
\frac{g_{\Sigma^* \pi \Sigma/\Lambda}}{m_{\pi}}\bar{\Sigma}^{*\mu}
(\partial_{\mu} \vec\tau \cdot \vec\pi) \Sigma/\Lambda \,+{\rm
h.c.}, \label{eq:pisigmal}
\end{eqnarray}
for the $u$-channel $\Sigma^0(1193)$ and $\Lambda(1115)$ exchange
diagrams.

The above Lagrangian densities are also used to study the
contributions of the terms shown in Fig.~\ref{ppdiagram} for $p p
\to n K^+ \Sigma^+(1385)$ reaction. In addition, we also need the
Lagrangian density as following for the $\pi NN$ vertex,
\begin{eqnarray}
{\mathcal L}_{\pi N N}  &=& -i g_{\pi NN} \bar{N} \gamma_5 \vec\tau
\cdot \vec\pi N .  \label{pinn}
\end{eqnarray}

\subsection{Coupling constants and form factors} \label{ccff}

Firstly, the coupling constant for $\pi NN$ vertex is taken to be
$g_{\pi NN}=13.45$, while the coupling constants $g_{KN\Sigma}$,
$g_{KN\Lambda}$, and $g_{K^* N \Sigma^*}$\footnote{In principle,
there are three terms for $K^* N \Sigma^*$ vertex as used in
Ref.~\cite{Oh:2007jd} (see Eq. (6) of that reference for more
details). However, there is no more information about this vertex,
and it is found that the other two couplings give minor
contributions to the $\gamma p \to K^+ \Sigma^0(1385)$
reaction~\cite{Oh:2007jd}. Thus, we ignore the contributions from
the other two couplings.} are respectively taken as $2.69$,
$-13.98$, and $-5.48$, which are obtained from the ${\rm SU}(3)$
flavor symmetry. And these values have also been used in previous
works~\cite{Doring:2010ap,Oh:2007jd,Kim:2012pz,Xie:2013wfa,Xie:2013db}.

Secondly, the coupling constants, $g_{K^*K\pi}$, $g_{\Sigma^* \pi
\Sigma}$, and $g_{\Sigma^*\pi \Lambda}$, are determined from the
experimentally observed partial decay widths of the $K^* \to K\pi$,
$\Sigma(1385) \to \pi \Sigma$, and $\Sigma(1385) \to \pi \Lambda$,
respectively. With the effective interaction Lagrangians described
by Eq.~(\ref{eq:kstarkpi}) and Eq.~(\ref{eq:pisigmal}), the partial
decay widths $\Gamma_{K^* \to K \pi}$ and $\Gamma_{\Sigma(1385) \to
\pi \Sigma/\Lambda}$ can be easily calculated. The coupling
constants are related to the partial decay widths as,

\begin{eqnarray}
\Gamma_{K^* \to K \pi} \! \!\! &=& \! \! \! \frac{g^2_{K^*K\pi}}{2\pi }
\frac{|\overrightarrow{p}^{{\rm c.m.}}_{\pi}|^3}{m^2_{K^*}}, \label{kstarkpi} \\
\Gamma_{\Sigma^* \to \pi \Sigma/\Lambda} \! \! \! &=& \! \! \!
\frac{ f_I g^2_{\Sigma^*\pi\Sigma/\Lambda}}{12 \pi}
\frac{|\overrightarrow{p}^{{\rm
c.m.}}_{\Sigma/\Lambda}|^3(E_{\Sigma/\Lambda} +
m_{\Sigma/\Lambda})}{m^2_{\pi} M_{\Sigma^*}}, \label{1385pisigma}
\end{eqnarray}
with the isospin factor $f_I = 2$ for $\Sigma^* \to \pi \Sigma$ and
$f_I =1$ for $\Sigma^* \to \pi \Lambda$, and
\begin{eqnarray}
E_{\Sigma/\Lambda}  &= & \frac{M^2_{\Sigma^*} + m^2_{\Sigma/\Lambda}
- m^2_{\pi}}{2M_{\Sigma^*}}, \nonumber \\
|\overrightarrow{p}^{{\rm c.m.}}_{\Sigma/\Lambda}| &= &
\sqrt{E^2_{\Sigma/\Lambda} - m^2_{\Sigma/\Lambda}}, \nonumber \\
|\overrightarrow{p}^{{\rm c.m.}}_{\pi}| &= & \frac{\sqrt{[m^2_{K^*}
- (m_K  +  m_{\pi})^2][m^2_{K^*} -  (m_K  - m_{\pi})^2]}}{2m_{K^*}}.
\nonumber
\end{eqnarray}

With mass ($M_{\Sigma^*} = 1384.57$ MeV, $m_{K^*} = 893.1$ MeV),
total decay width ($\Gamma_{\Sigma^*} = 37.13$ MeV, $\Gamma_{K^*} =
49.3$ MeV), and decay branching ratios of $\Sigma(1385)$
[Br($\Sigma^* \to \pi \Sigma$) = $0.117 \pm 0.015$, Br($\Sigma^* \to
\pi \Lambda$) = $0.87 \pm 0.015$] and $K^*$ [Br($K^* \to K\pi$)
$\sim 1$], we obtain these coupling constants as listed in
Table~\ref{tab1}.

\begin{table}[htbp]
\begin{center}
\caption{\label{table} Values of the coupling constants required for
the estimation of the $\pi^+ p \to K^+ \Sigma^+(1385)$ and $p p \to
n K^+ \Sigma^+(1385)$ reactions. These have been estimated from the
decay branching ratios quoted in the PDG book~\cite{pdg2012}, though
it should be noted that these are for all final charged state.}
\begin{tabular}{|ccc|}
\hline Decay modes  & Adopted branching ratios & $g$~\footnote{It
should be stressed that the partial decay width determine only the
square of the corresponding coupling constants as shown in
Eqs.~(\ref{kstarkpi}, \ref{1385pisigma}), thus their signs remain
uncertain. Predictions from quark model can be used to constrain
these signs. Unfortunately, quark model calculations for these
vertices are still sparse. We thus choose a positive sign for these coupling constants.}\\
\hline $\Sigma^* \to \pi \Lambda$  & $0.87$ & $1.26$ \\
       $\Sigma^* \to \pi \Sigma$  & $0.12$ & $0.69$ \\
       $K^* \to K\pi$      & $1.00$  & $3.24$ \\
\hline
\end{tabular} \label{tab1}
\end{center}
\end{table}

Finally, the strong coupling constants $g_{\pi N \Delta^*}$ and
$g_{1,2}$ for the $\Delta^*(1940) \Sigma(1385) K$ vertex are free
parameters, which will be determined by fitting to the experimental
data on the total cross sections of the $\pi^+ p \to K^+
\Sigma^+(1385)$ reaction.

In evaluating the scattering amplitudes of $\pi^+ p \to K^+
\Sigma^+(1385)$ and $pp \to n K^+ \Sigma^+(1385)$ reactions, we need
to include the form factors because the hadrons are not point like
particles. We adopt here the common scheme used in many previous
works,
\begin{eqnarray}
&& f_i =\frac{\Lambda^4_i}{\Lambda^4_i+(q_i^2-M_i^2)^2},
\quad i= s, t, u, \\
&&\quad {\rm with} ~~~~~~ \left\{\begin{array}{l}  q_s^2=s, \,
q_t^2= t, \, q_u^2= u, \cr M_s = M_{\Delta^*}, M_t = m_{K^*}, \cr
M_u = m_{\Sigma}, m_{\Lambda},
\end{array}\right. \label{pipff}
\end{eqnarray}
where $s$, $t$ and $u$ are the Lorentz-invariant Mandelstam
variables. In the present calculation, $q_s=p_1+p_2$, $q_t=p_1-p_3$,
and $q_u=p_4-p_1$ are the 4-momentum of intermediate
$\Delta^*(1940)$ resonance in the $s$-channel, exchanged
$K^{*0}(892)$ meson in the $t$-channel, and exchanged
$\Sigma^0(1193)$ and $\Lambda(1115)$ in the $u$-channel,
respectively. While $p_1,~ p_2,~ p_3$ and $p_4$ are the 4-momenta
for $\pi^+$, $p$, $K^+$ and $\Sigma^+(1385)$, respectively. In
principle, the cutoff $\Lambda_s$, $\Lambda_t$ and $\Lambda_u$ are
free parameters of the model, but in practice we will constrain them
to a common value between $0.6$ and $1.2$ GeV. By doing this, we can
reduce the number of the free parameters.

\subsection{Scattering amplitudes}

The invariant scattering amplitudes that enter our model for
calculation of the total cross sections for the
\begin{eqnarray}
\pi^+ (p_1) p(p_2,s_p) \to K^+ (p_3)
\Sigma^+(1385)(p_4,s_{\Sigma^*})
\end{eqnarray}
are defined as
\begin{equation}
-iT_i=\bar u_\mu(p_4,s_{\Sigma^*}) A_i^{\mu} u(p_2,s_p), \label{ti}
\end{equation}
where $u_\mu$ and $u$ are dimensionless Rarita-Schwinger and Dirac
spinors, respectively, while $s_{\Sigma^*}$ and $s_p$ are the spin
polarization variables for final $\Sigma^+(1385)$ and initial
proton, respectively. To get the scattering amplitudes, we need also
the propagators for $\Delta^*(1940)$, $K^*$ meson, and
$\Sigma^0/\Lambda$ hyperon,
\begin{eqnarray}
G^{\mu\nu}_{K^*}(q_t) &=& i \frac{-g^{\mu\nu}+q^{\mu}_t q^{\nu}_t/m^2_{K^*} }{t-m^2_{K^*}}, \\
G_{\Sigma/\Lambda}(q_u) &=& i \frac{\Slash q_u + m_{\Sigma/\Lambda}}{u-m^2_{\Sigma/\Lambda}}, \\
G^{\mu\nu}_{\Delta^*}(q_s) &=& i \frac{\Slash q_s +
M_{\Delta^*}}{D}P^{\mu\nu},
\end{eqnarray}
with
\begin{eqnarray}
D &=& s-M^2_{\Delta^*} + iM_{\Delta^*}\Gamma_{\Delta^*}, \\
P^{\mu \nu} &=& -g^{\mu \nu} + \frac{1}{3}\gamma^{\mu}\gamma^{\nu} +
\frac{2}{3M^2_{\Delta^*}}q_s^{\mu}q_s^{\nu} \nonumber \\
&& +
\frac{1}{3M_{\Delta^*}}(\gamma^{\mu}q_s^{\nu}-\gamma^{\nu}q_s^{\mu}),
\end{eqnarray}
where $M_{\Delta^*}$ and $\Gamma_{\Delta^*}$ are the mass and total
decay width of the $\Delta^*(1940)$ resonance, respectively. Because
$M_{\Delta^*}$ and $\Gamma_{\Delta^*}$ have large experimental
uncertainties~\cite{pdg2012}, we take them as free parameters and
will fit them to the total cross sections of $\pi^+ p \to K^+
\Sigma^+(1385)$ reaction.

Then, the reduced  $A_i^{\mu}$ amplitudes in Eq.~(\ref{ti}) can be easily obtained,
\begin{eqnarray}
A_{s}^{\mu} &=& i \frac{g_{\pi N \Delta^*}}{m_{\pi}D} \Big [
\frac{g_1}{m_K} \Slash p_3 (\Slash q_s + M_{\Delta^*})
\Big( p^{\mu}_1 - \frac{1}{3}\gamma^{\mu} \Slash p_1 - \nonumber \\
&& \frac{1}{3M_{\Delta^*}}(\gamma^{\mu}q_s \cdot p_1 - q^{\mu}_s \Slash p_1) -
\frac{2}{3M^2_{\Delta^*}}q^{\mu}_s q_s \cdot p_1 \Big) - \nonumber \\
&& \frac{g_2}{m^2_K} (\Slash q_s + M_{\Delta^*})  p^{\mu}_3
\Big( p_1 \cdot p_3 - \frac{1}{3} \Slash p_3 \Slash p_1 - \nonumber \\
&& \frac{1}{3M_{\Delta^*}}(\Slash p_3 q_s \cdot p_1 - q_s \cdot p_3 \Slash p_1) - \nonumber \\
&& \frac{2}{3M^2_{\Delta^*}} q_s \cdot p_3 q_s \cdot p_1 \Big) \Big] f_s, \label{eq:as} \\
A_t^{\mu} &=& \frac{\sqrt{2} g_{K^* K\pi}g_{K^* N
\Sigma^*}}{m_N(t-m^2_{K^*})}
(\Slash p_3 p^{\mu}_1 - \Slash p_1 p^{\mu}_3) f_t, \, \label{eq:at} \\
A_u^{\mu} &=& i \frac{g_{\Sigma^* \pi \Sigma/\Lambda}g_{KN
\Sigma/\Lambda}}{m_{\pi}(u - m^2_{\Sigma/\Lambda})}(\Slash q_u +
m_{\Sigma/\Lambda})  \gamma_5 p_1^{\mu}\, f_u, \label{eq:au}
\end{eqnarray}
with the sub-indices $s$, $t$ and $u$ stand for the $s$-channel
$\Delta^*(1940)$, $t$-channel $K^{*0}(892)$ exchange, and
$u$-channel $\Sigma^0(1193)$ and $\Lambda(1115)$ exchange,
respectively. As we can see, in the tree-level approximation, only
the products, such as $ g_1g_{\pi N \Delta^*}$ ($\equiv
\tilde{g}_1$) and $ g_2 g_{\pi N \Delta^*}$ ($\equiv \tilde{g}_2$)
enter the invariant scattering amplitudes. Because the information
on these couplings are scarce, they are also determined by fitting
them to the low-energy experimental data on the total cross sections
of $\pi^+ p \to K^+ \Sigma^+(1385)$ reaction.

For the $pp \to n K^+ \Sigma^+(1385)$ reaction, the full invariant
scattering amplitude in our calculation is composed of two parts
corresponding to the $s$-channel $\Delta^*(1940)$ resonance, and
$u$-channel $\Lambda(1115)$ hyperon pole, which are produced by the
$\pi^+$-meson exchanges,
\begin{eqnarray}
{\cal M} = \sum_{i = s,~u} {\cal M}_{i}. \label{ppamp}
\end{eqnarray}

Each of the above amplitudes can be obtained straightforwardly with
the effective couplings and following the Feynman rules. Here we
give explicitly the amplitude ${\cal M}_s$, as an example,
\begin{eqnarray}
{\cal M}_s & = & \frac{\sqrt{2} g_{\pi NN} g_{\pi N
\Delta^*}}{m_{\pi}} F^{N N}_{\pi}(k^2_{\pi}) F^{\Delta^*
N}_{\pi}(k^2_{\pi}) F_s(q_{\Delta^*}^2)\times \nonumber\\
&&  G_{\pi}(k^2_{\pi}) \bar{u}^{\mu} (p_4,s_4)
(-\frac{g_1}{m_K} \Slash p_5 g_{\mu \rho} + \frac{g_2}{m^2_K}p_{5 \mu} p_{5\rho}) \times  \nonumber\\
&&  G^{\rho \sigma}_{\Delta^*}(q_s) k_{\pi \sigma}
\gamma_5 u(p_1,s_1)\bar{u}(p_3,s_3) \gamma_5 u(p_2,s_2)  \nonumber\\
&&  + (\text {exchange term with } p_1 \leftrightarrow p_2),
\label{ppampms}
\end{eqnarray}
where $s_i~(i=1,2,3)$ and $p_i~(i=1,2,3)$ represent the spin
projection and 4-momenta of the two initial protons and final
neutron, respectively. While $p_4$ and $p_5$ are the 4-momenta of
the final $\Sigma^+(1385)$ and $K^+$ meson, respectively. And $s_4$
stands for the spin projection of $\Sigma^+(1385)$. In
Eq.~(\ref{ppampms}), $k_{\pi} = p_2 - p_3$ and $q_{\Delta^*} = p_4 +
p_5$ stand for the 4-momenta of the exchanged $\pi^+$ meson and
intermediate $\Delta^{*}(1940)$ resonance. And $G_{\pi}(k^2_{\pi})$
is the pion meson propagator,
\begin{equation}
G_{\pi}(k^2_{\pi})=\frac{i}{k_{\pi}^2-m^2_{\pi}}.
\end{equation}

For $p p \to p K^+ \Lambda(1520)$ reaction, we need also the
relevant off-shell form factors for $\pi NN$ and $\pi N \Delta^*$
vertexes, which have been already included in the amplitude of
Eq.~(\ref{ppampms}), and we take them as,
\begin{eqnarray}
F^{NN}_{\pi}(k^2_{\pi}) &=& \frac{\Lambda^2_{\pi}-m_{\pi}^2}
{\Lambda^2_{\pi}- k_{\pi}^2}, \label{pinnff} \\
F^{\Delta^* N}_{\pi}(k^2_{\pi}) &=&
\frac{\Lambda^{*2}_{\pi}-m_{\pi}^2}{\Lambda^{*2}_{\pi}-k_{\pi}^2},
\label{pinnstarff}
\end{eqnarray}
with $k_{\pi}$ the 4-momentum of the exchanged $\pi$ meson. The
cutoff parameters are taken as $\Lambda_{\pi} = \Lambda^*_{\pi}$ =
1.1 GeV, with which the experimental data on $p p \to n K^+
\Sigma^+(1385)$ reaction can be reproduced.

\subsection{Cross sections for $\pi^+ p \to K^+ \Sigma^+(1385)$ reaction}

The differential cross section for $\pi^+ p \to K^+ \Sigma^+(1385)$
reaction at center of mass ($\rm c.m.$) frame can be expressed as
\begin{equation}
{d\sigma \over d{\rm cos}\theta}={1\over 32\pi s}{
|\vec{p}_3^{\text{~c.m.}}| \over |\vec{p}_1^{\text{~c.m.}}|} \left (
{1\over 2}\sum_{s_{\Sigma^*},s_p}|T|^2 \right ), \label{eq:pipdcs}
\end{equation}
where $\theta$ denotes the angle of the outgoing $K^+$ relative to
beam direction in the $\rm c.m.$ frame, while
$\vec{p}_1^{\text{~c.m.}}$ and $\vec{p}_3^{\text{~c.m.}}$ are the
3-momentum of the initial $\pi^+$ and final $K^+$ mesons. The total
invariant scattering amplitude $T$ is given by,
\begin{equation}
T=T_s + T_t + T_u \, .
\end{equation}

From the amplitude, we can easily obtain the total cross sections of
the $\pi^+ p \to K^+ \Sigma^+(1385)$ reaction as functions of the
beam momentum $p_{\pi^+}$. By including all the contributions from
the $s$-channel $\Delta^*$ resonance, $t$-channel $K^{*0}(892)$, and
$u$-channel $\Sigma^0(1193)$ and $\Lambda(1115)$ processes, at fixed
cutoff parameters $\Lambda_s \neq \Lambda_t = \Lambda_u$, we perform
four parameter ($M_{\Delta^*}$, $\Gamma_{\Delta^*}$, $\tilde{g}_1$,
and $\tilde{g}_2$) $\chi^2-$fit to the experimental data on total
cross sections for $\pi^+ p \to K^+ \Sigma^+(1385)$ reaction. There
is a total of $11$ data points below $p_{\pi^+} = 4$ GeV.

\begin{figure}[htbp]
\begin{center}
\includegraphics[scale=0.4]{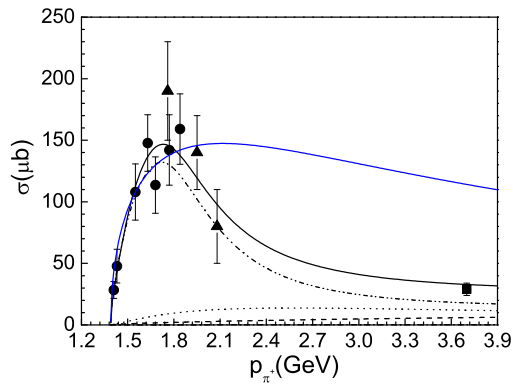}
\caption{(Color online) Total cross sections vs the beam momentum
$p_{\pi^+}$ for $\pi^+ p \to K^+ \Sigma^+(1385)$ reaction. The
experimental data are taken from Ref.~\cite{Hanson:1972zza} (dots),
Ref.~\cite{dagan} (triangles), and Ref.~\cite{Butler:1973gq}
(square). The curves are the contributions from $s$-channel
$\Delta^*(1940)$ (dash-dot-dotted), $t$-channel $K^{*0}$ (dashed),
$u$-channel $\Sigma^0(1193)$ (dash-dotted) and $\Lambda(1115)$
(dotted), and the total contributions of them (black-solid),
respectively. The blue-solid cure is obtained from the
Stodolsky-Sakural model which will be discussed below.}
\label{piptcs}
\end{center}
\end{figure}

By constraining the value of the cutoff parameters $\Lambda_s$ and
$\Lambda_t = \Lambda_u$ from $0.6$ to $1.2$ GeV, we get the minimal
$\chi^2/dof$ $0.8$ with $ \Lambda_t = \Lambda_u = 0.6$ GeV and
$\Lambda_s=0.9$ GeV, and the fitted parameters are: $M_{\Delta^*} =
1940 \pm 24$ MeV, $\Gamma_{\Delta^*} = 172 \pm 94$ MeV, $\tilde{g}_1
= -0.36 \pm 0.19$, and $\tilde{g}_2 = 1.83 \pm 0.16$. The best
fitting results for the total cross sections are shown in
Fig.~\ref{piptcs}, comparing with the experimental data from
Refs.~\cite{Hanson:1972zza,dagan,Butler:1973gq}. The black-solid
line represents the full results, while the contributions from the
$s$-channel $\Delta^{++*}(1940)$ resonance, $t$-channel
$K^{*0}(892)$ exchange, $u$-channel $\Lambda(1115)$ and
$\Sigma^0(1193)$ terms are shown by the dash-dot-doted, dashed,
dotted, and dash-doted lines, respectively. From Fig.~\ref{piptcs},
one can see that the description of the experimental data is quite
well, especially, thanks to the contributions from the
$\Delta^*(1940)$ resonance, the bump structure around $p_{\pi^+} =
1.8$ GeV can be described well. It is also show that the $s$-channel
$\Delta^*(1940)$ resonance gives the dominant contribution, while
the $t$-channel and $u$-channel diagrams give the minor
contributions.

\begin{figure*}[htbp]
\begin{center}
\includegraphics[scale=0.4]{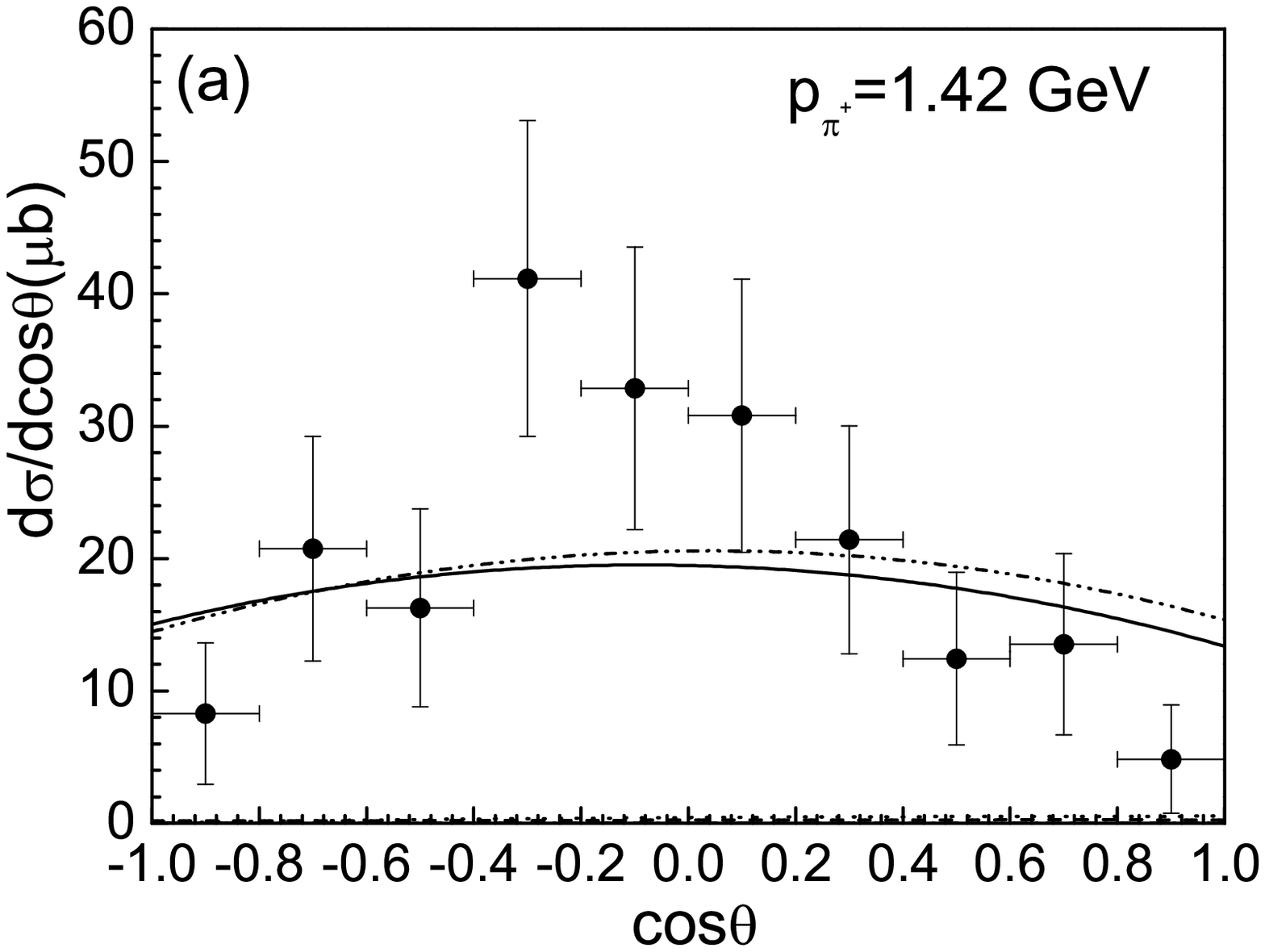}
\includegraphics[scale=0.4]{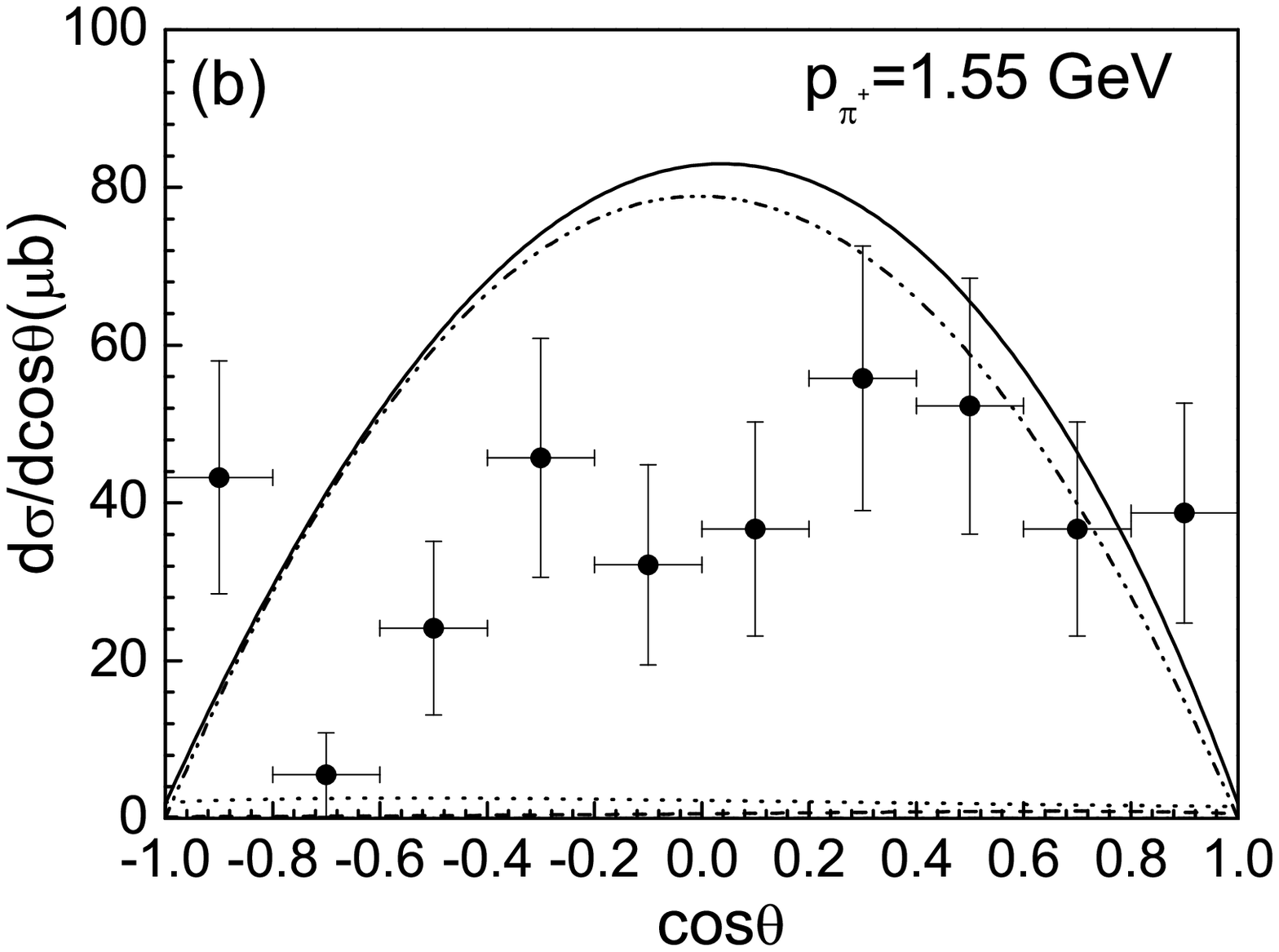}
\includegraphics[scale=0.4]{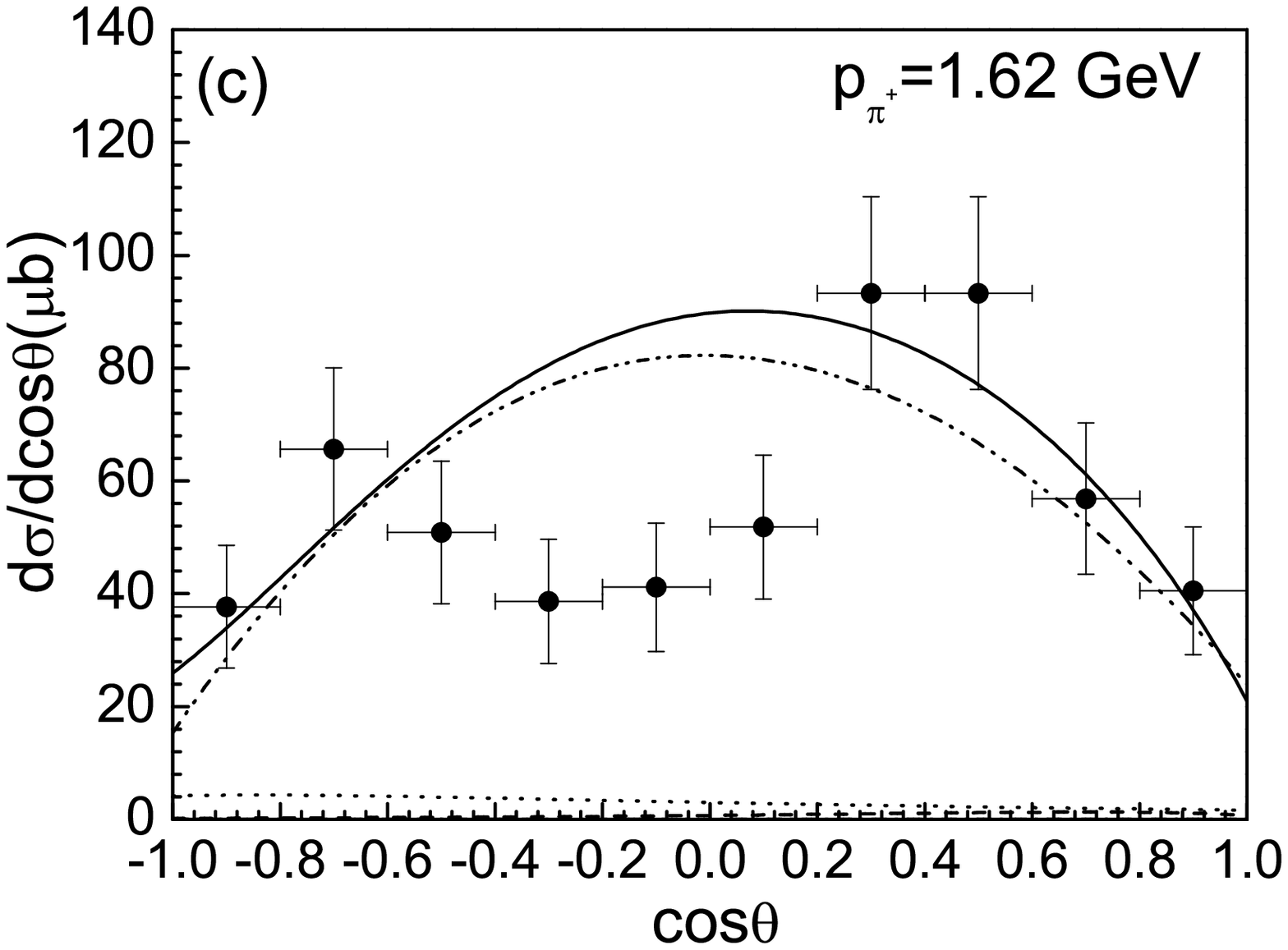}
\includegraphics[scale=0.4]{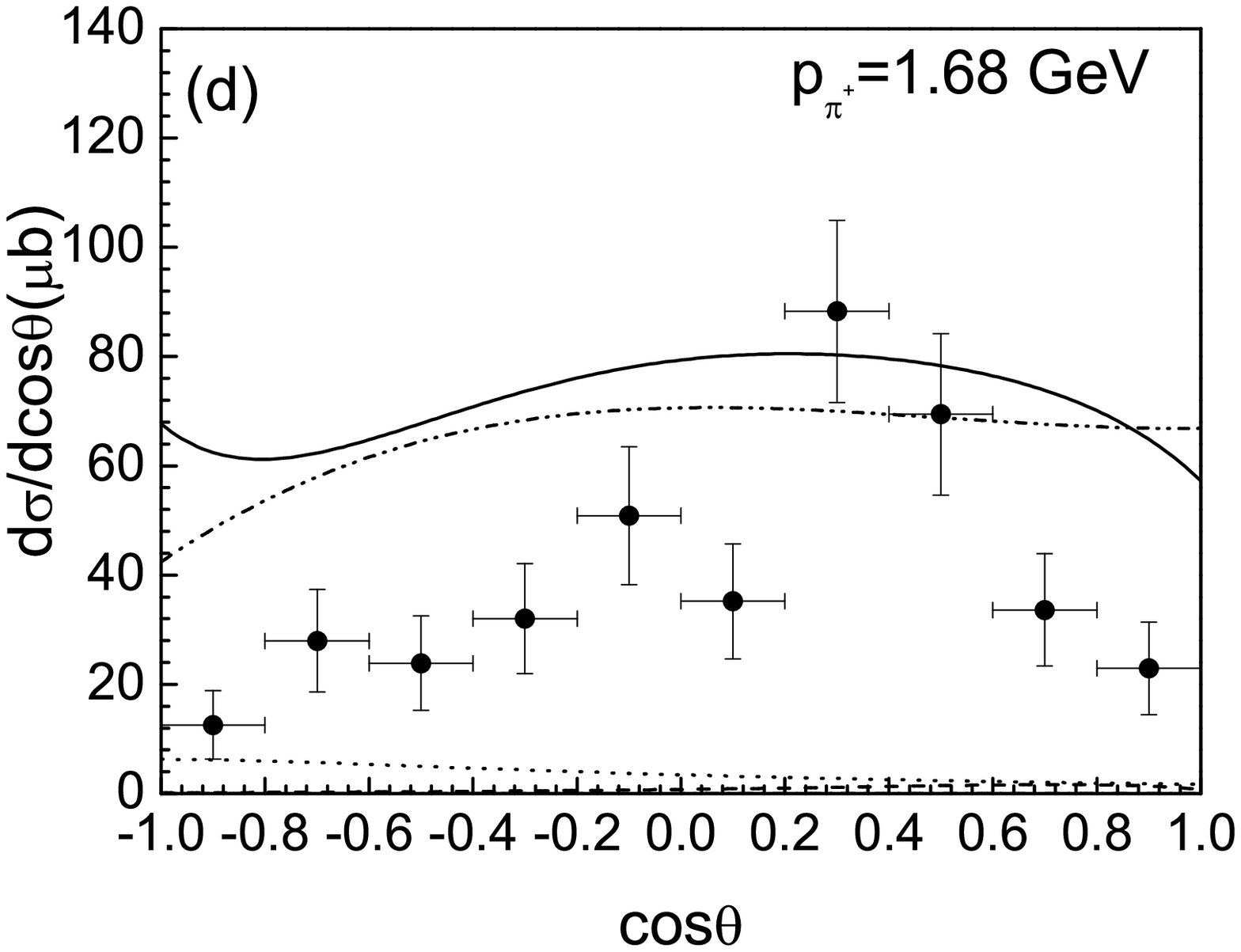}
\includegraphics[scale=0.4]{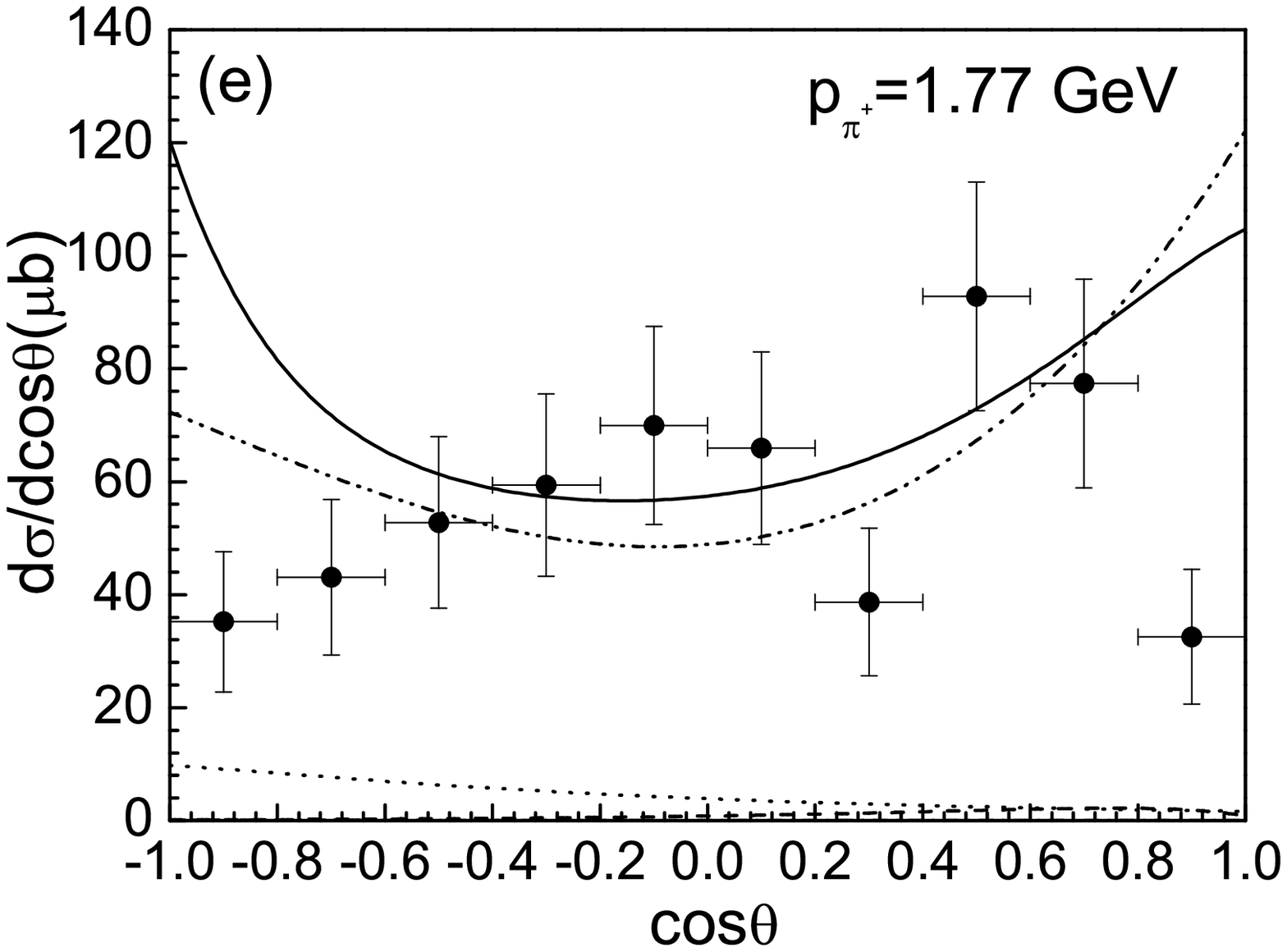}
\includegraphics[scale=0.4]{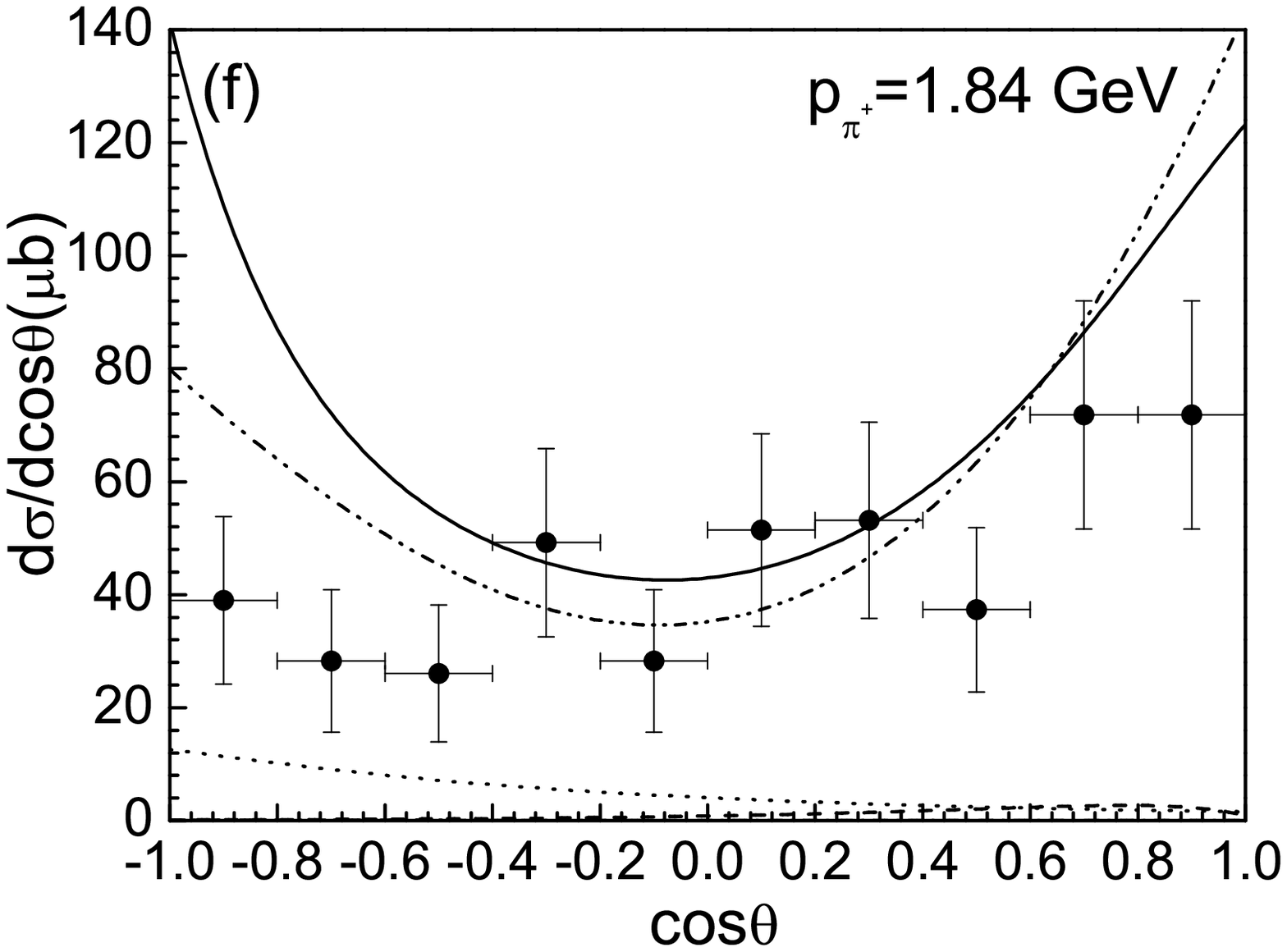}
\caption{Predictions of the differential cross sections,
$d\sigma/d{\rm cos}\theta$, for $\pi^+ p \to K^+ \Sigma^+(1385)$
reaction at different beam momentum. The experimental data are taken
from Ref.~\cite{Hanson:1972zza}. The curves are the contributions
from $s$-channel $\Delta^*(1940)$ (dash-dot-dotted), $t$-channel
$K^{*0}$ (dashed), $u$-channel $\Sigma^0(1193)$ (dash-dotted) and
$\Lambda(1115)$ (dotted), and the total contributions of them
(solid), respectively.} \label{pipdcs}
\end{center}
\end{figure*}

In Fig.~\ref{pipdcs}, the corresponding model predictions for the
differential cross sections, $d\sigma/d{\rm cos}\theta$, of the
$\pi^+ p \to K^+ \Sigma^+(1385)$ reaction are shown. Those results
are obtained at $p_{\pi^+} =1.42$ GeV [Fig.~\ref{pipdcs}(a)],
$p_{\pi^+} =1.55$ GeV [Fig.~\ref{pipdcs}(b)], $p_{\pi^+} =1.62$ GeV
[Fig.~\ref{pipdcs}(c)], $p_{\pi^+} =1.68$ GeV
[Fig.~\ref{pipdcs}(d)], $p_{\pi^+} =1.77$ GeV
[Fig.~\ref{pipdcs}(e)], and $p_{\pi^+} =1.84$ GeV
[Fig.~\ref{pipdcs}(f)], respectively. We also show the experimental
data taken from Ref.~\cite{Hanson:1972zza} for comparison. One can
see that by considering the dominant contributions from the
$\Delta^*(1940)$, our model calculations can reasonably describe the
angular distributions within the large experimental errors. However,
at some energy points, such as $p_{\pi^+} =1.55$ GeV
[Fig.~\ref{pipdcs}(b)], $p_{\pi^+} =1.62$ GeV
[Fig.~\ref{pipdcs}(c)], and $p_{\pi^+} =1.68$ GeV
[Fig.~\ref{pipdcs}(d)], our model calculations can not well
reproduce the experimental measurements.

It is pointed out that the Stodolsky-Sakural
model~\cite{Stodolsky:1963kt,Stodolsky:1964zz} with dominant
contribution from $t$-channel $K^*$ exchange fits those production
angular distributions reasonably well at all beam
momenta~\cite{Hanson:1972zza} (see more details in Fig. 4 of that
reference). The predictions of this model are that the form of the
differential cross sections for $\pi^+ p \to K^+ \Sigma^+(1385)$
reaction is given by~\cite{Hanson:1972zza}
\begin{equation}
\frac{d\sigma} {d {\rm cos}\theta} \propto \frac{1-{\rm
cos}^2\theta}{(t-M^2_{K^*})^2},
\end{equation}
from where we can obtain the total cross sections~\footnote{We
include also the phase space factor, $|\vec{p}_3^{\rm{~c.m.}}|$, in
our estimation. In this way the total cross section is obtained from
$\sigma = N \int^{1}_{-1} \frac{1-{\rm
cos}^2\theta}{(t-M^2_{K^*})^2} |\vec{p}_3^{\rm{~c.m.}}| d{\rm
cos}\theta$, with a normalization $N = 1.54$ GeV.} as shown in
Fig.~\ref{piptcs} by the blue-solid curve. One can see that the
$t$-channel $K^*$ exchange can reproduce well the experimental data
from Ref.~\cite{Hanson:1972zza}, but, it can not give the bump
structure if we take those measurements of
Refs.~\cite{dagan,Butler:1973gq} into account as shown in
Fig.~\ref{piptcs}. Thus, the Stodolsky-Sakural model can reasonably
describe the angular distribution at all momenta should not be
surprising since the it considered only the experimental data from
Ref.~\cite{Hanson:1972zza}, where the bump structure does not appear
because of the narrow energy range of measurements of
Ref.~\cite{Hanson:1972zza}.

On the other hand, we find that the experimental results of
differential cross sections of Ref.~\cite{Hanson:1972zza} and the
total cross sections data of
Refs.~\cite{Hanson:1972zza,dagan,Butler:1973gq} can not be
simultaneously fitted well, which is because the differential cross
sections data with large uncertainties are inconsistent between
different angles and energies, hence, those data points about the
differential cross sections from Ref.~\cite{Hanson:1972zza} are not
taken into account in our best fit.

\subsection{Partial decay widths of $\Delta^*(1940)$ resonance}

With the Lagrangian densities of Eqs.~(\ref{pinnstar}) and
(\ref{eq:eqknstar}), we can evaluated the $\Delta^*(1940)$ to $N
\pi$ and $\Delta^*(1940)$ to $\Sigma(1385) K$ partial decay widths,
\begin{eqnarray}
&& \Gamma_{\Delta^* \to  N \pi} = \frac{g^2_{\pi N \Delta^*}}{12\pi}
\frac{|\vec{p}_N^{\,\,\rm
c.m.}|^3}{m^2_{\pi}M_{\Delta^*}}(E_N-m_N), \label{pdwpin} \\
&& \Gamma_{\Delta^* \to  \Sigma^* K } = \frac{|\vec{p}_1^{\,\,\rm
cm}| (E_{\Sigma^*} + M_{\Sigma^*})}{36 \pi M_{\Delta^*}
M_{\Sigma^*}^2} \Big \{
 \frac{2g_2^2M^2_{\Delta^*}}{m^4_K}  |\vec{p}_1^{\,\,\rm cm}|^4 + \nonumber \\
& & \frac{2g_1g_2}{m_K^3} M_{\Delta^*} (M_{\Delta^*} - M_{\Sigma^*})
(2 E_{\Sigma^*} + M_{\Sigma^*}) |\vec{p}_1^{\,\,\rm cm}|^2 +
\nonumber \\
 &&  \frac{g_1^2}{m_K^2} ( M_{\Delta^*} \! - \! M_{\Sigma^*})^2 ( 2E_{\Sigma^*}^2 \! + \!
  2E_{\Sigma^*}M_{\Sigma^*} \! + \! 5 M_{\Sigma^*}^2 )
  \Big \} , \label{pdwksigma}
\end{eqnarray}
where,
\begin{eqnarray}
E_N & =& \frac{M^2_{\Delta^*}+m^2_N-m^2_{\pi}}{2M_{\Delta^*}}, \\
|\vec{p}_N^{\,\,\rm c.m.}| &=& \sqrt{E^2_N-m^2_N}, \\
E_{\Sigma^*} &=& \frac{M^2_{\Delta^*} + M^2_{\Sigma^*} - m^2_K }{2M_{\Delta^*}}, \\
|\vec{p}_1^{\,\,\rm cm}| &=& \sqrt{E^2_{\Sigma^*} - M_{\Sigma^*}^2}
\, .
\end{eqnarray}

With the values of $M_{\Delta^*}$, $\Gamma_{\Delta^*}$,
$\tilde{g}_1$ and $\tilde{g}_2$ obtained from the present fit, we
get Br($\Delta^* \to  N \pi$)$\times$Br($\Delta^* \to \Sigma^* K $)
$=(0.52 \pm 0.13)\%$ with the error from the uncertainty of the
fitted parameters.

On the other hand, the fitted results for the mass and total decay
width of the $\Delta^*(1940)$ resonance are compatible with previous
analysis in Ref.~\cite{cutkosky},
\begin{eqnarray}
M_{\Delta^*(1940)} &=& 1940 \pm 100 ~{\rm MeV} ,\\
\Gamma_{\Delta^*(1940)} &=& 200 \pm 100 ~{\rm MeV},
\end{eqnarray}
quoted in PDG~\cite{pdg2012}. Next, by using the branch ratio of
Br($\Delta^*(1940) \to N\pi$) obtained in Ref.~\cite{cutkosky} and
the total decay width of $\Gamma_{\Delta^*(1940)}$ from our present
fit, we can determine the strong coupling constant, $g_{\pi N
\Delta^* } = 0.35 \pm 0.12$ from the relation of Eq.~(\ref{pdwpin}).
Then we can easily obtain the values of the strong $\Delta^*(1940)
\Sigma(1385) K$ coupling constants $g_1$ and $g_2$,
\begin{eqnarray}
g_1 &=& -1.04 \pm 0.38 ,\\
g_2 &=& 5.24 \pm 2.30 .
\end{eqnarray}

Furthermore, the branch ration Br($\Delta^* \to \Sigma^* K$) and
partial decay width $\Gamma_{\Delta^* \to \Sigma^* K}$ are $(10.4
\pm 4.9)\%$ and $17.9 \pm 12.9$ MeV, respectively. We find that the
$\Sigma^* K$ decay mode of the $\Delta^*(1940)$ resonance could be
larger than the $N \pi$ channel, if one attributes the bump
structure in the total cross sections of $\pi^+ p \to K^+
\Sigma^+(1385)$ reaction~\cite{Hanson:1972zza,dagan,Butler:1973gq},
to the effects produced by this resonance, as implicitly assumed in
this work. This large coupling of the two-star $D-$wave $J^P=3/2^-$
$\Delta^*(1940)$ resonance to the $\Sigma^* K^+$ channel will
confirm/get support from the QM results of Capstick, and Roberts in
Ref.~\cite{simonprd58}, as mentioned above.

\section{Numerical results for $pp \to n K^+ \Sigma^+(1385)$ reaction}

With the formalism and ingredients given above, the calculations of
the differential and total cross sections for $pp \to n K^+
\Sigma^+(1385)$ are straightforward,
\begin{eqnarray}
&& d\sigma (pp \to n K^+ \Sigma^+(1385)) =
\frac{1}{4}\frac{m^2_p}{F}
\sum_{s_1, s_2} \sum_{s_3, s_4} |{\cal M}|^2 \times \nonumber \\
&& \frac{m_n d^{3} p_{3}}{E_{3}}
\frac{m_{\Sigma^+(1385)} d^{3} p_4}{E_4} \frac{d^{3} p_5}{2 E_5} \delta^4 (p_{1}+p_{2}-p_{3}-p_{4}-p_5), \nonumber \\
\label{ppdcs}
\end{eqnarray}
with the flux factor
\begin{eqnarray}
F=(2 \pi)^5\sqrt{(p_1\cdot p_2)^2-m^4_p}~. \label{eqff}
\end{eqnarray}

\begin{figure}[htbp]
\begin{center}
\includegraphics[scale=0.4]{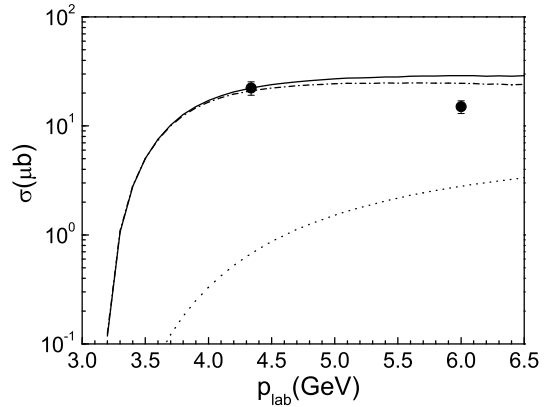}
\caption{Total cross sections vs beam energy ${\rm p_{lab}}$ of
proton for the $pp \to n K^+\Sigma^+(1385)$ reaction from present
calculation. The dotted and dash-dotted lines stand for
contributions from $\Lambda(1115)$ pole and $\Delta^*(1940)$
resonance, respectively. Their total contribution are shown by the
solid line. The experimental data are taken from
Refs.~\cite{Agakishiev:2011qw,Klein:1970ri}.} \label{pptcs}
\end{center}
\end{figure}

The total cross section versus the beam energy (${\rm p_{lab}}$) of
the proton for the $pp \to n K^+ \Sigma^+(1385)$ reaction is
calculated by using a Monte Carlo multi-particle phase space
integration program. The results for beam energies ${\rm p_{lab}}$
from just above the production threshold $3.2$ GeV to $6.5$ GeV are
shown in Fig.~\ref{pptcs}. The dotted, and dash-dotted lines stand
for contributions from $\Lambda(1115)$ and $\Delta^{*}(1940)$
resonance, respectively. Their total contributions are shown by the
solid line.~\footnote{Since the $t$-channel $K^{*0}$ meson and
$u$-channel $\Sigma^0(1193)$ exchange give very small contribution
to the $\pi^+ p \to K^+ \Sigma^+(1385)$ reaction, especially for the
invariant mass of $K\Sigma(1385)$ around $2$ GeV, we ignore these
contributions in the calculation for the $pp \to n
K^+\Sigma^+(1385)$ reaction.} From Fig.~\ref{pptcs}, we can see that
the contribution from the $\Delta^*(1940)$ resonance is predominant
in the whole considered energy region. For comparison, we also show
the experimental data~\cite{Agakishiev:2011qw,Klein:1970ri} in
Fig.~~\ref{pptcs}, from where we can see that our predictions for
the total cross sections of $pp \to n K^+ \Sigma^+(1385)$ reaction
are in agreement with the experimental measurements.

In addition to the total cross sections, we also compute the
differential distributions for $pp \to n K^+ \Sigma^+(1385)$
reaction, namely the angular distributions of all final-state
particles in the overall center-of-mass frame (CMS), as well as
distributions in both the Gottfried-Jackson and helicity frames as
introduced in Refs.~\cite{Agakishiev:2011qw,AbdelBary:2010pc}. Like
Dalitz plots, the helicity angle distributions provide insight into
the three-body final state. While the information contained in the
Gottfried-Jackson angle distributions is complementary to that of a
Dalitz plot, as this angular distribution can give insight into the
scattering process, especially concerning the involved partial
waves.

\begin{figure*}[htbp]
\includegraphics[scale=0.8]{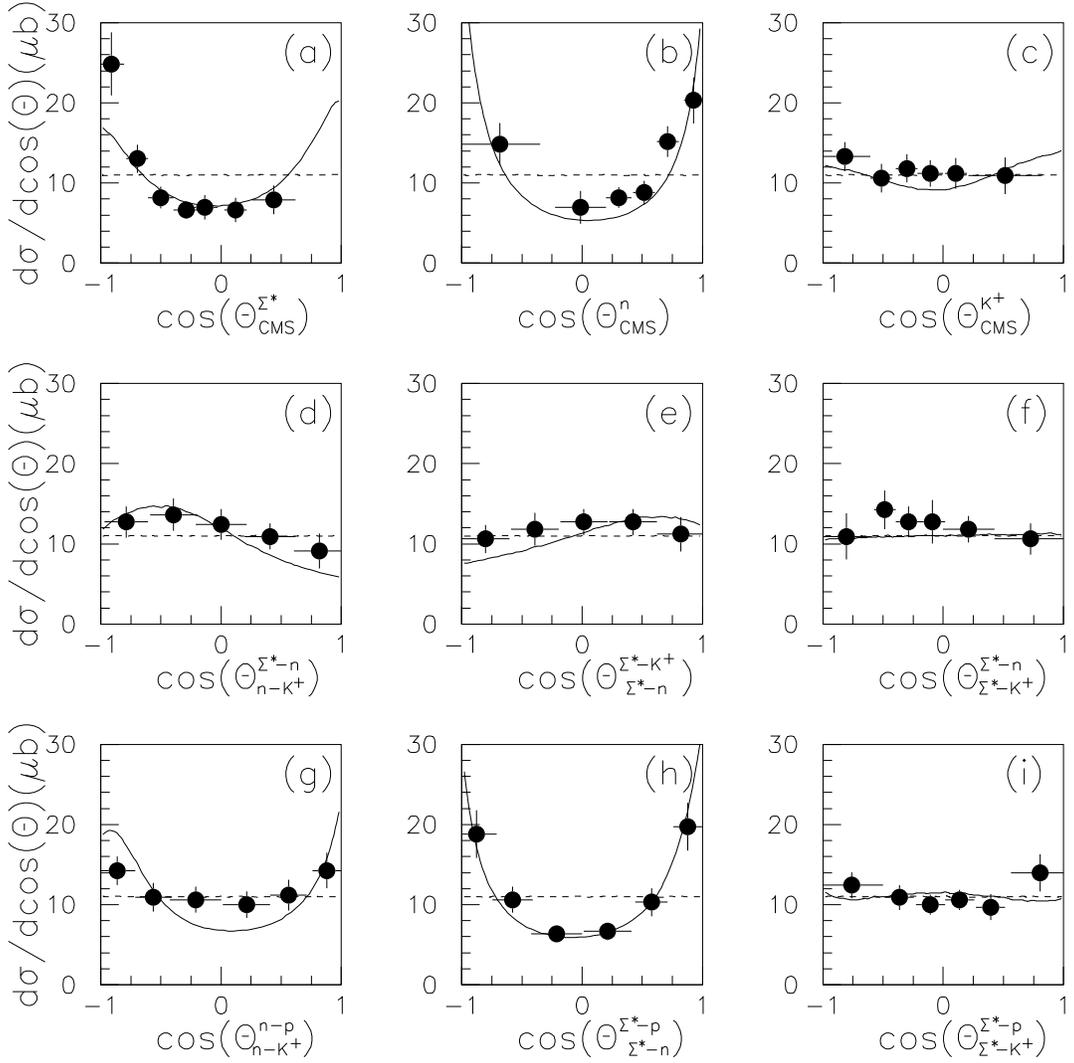}
\vspace{-0.2cm} \caption{Angular differential cross sections for the
$pp \to n K^+ \Sigma^+(1385)$ reaction in CMS [(a):
$\Theta^{\Sigma^*}_{\rm CMS}$, (b): $\Theta^n_{\rm CMS}$, (c):
$\Theta^{K^+}_{\rm CMS}$], helicity [(d):
$\Theta^{\Sigma^*-n}_{n-K^+}$, (e):
$\Theta^{\Sigma^*-K^+}_{\Sigma^*-n}$, (f):
$\Theta^{\Sigma^*-n}_{\Sigma^*-K^+}$], and Gottfried-Jackson [(g):
$\Theta^{n-p}_{n-K^+}$, (h): $\Theta^{\Sigma^*-p}_{\Sigma^*-n}$,
(i): $\Theta^{\Sigma^*-p}_{\Sigma^*-K^+}$] reference frames. The
dashed lines are pure phase space distributions, while the solid
lines are full results from our model. The experimental data are
taken from Ref.~\cite{Agakishiev:2011qw}.} \label{Fig:angdis}
\end{figure*}

The corresponding theoretical results are shown in
Fig.~\ref{Fig:angdis} with the experimental data taken from
Ref.~\cite{Agakishiev:2011qw}, where the dashed lines are pure phase
space distributions, while the solid lines are full results from our
model. We can see that our theoretical results with the dominant
contributions from the $\Delta^*(1940)$ resonance can describe the
experimental data fairly well, and only the phase space is far from
the data. The agreement of our model calculation with the
experimental data in Fig.~\ref{Fig:angdis} indicates that the HADES
data support the important role played by an odd-parity $3/2^-$
$\Delta^{*}(1940)$ resonance with a mass in the region of $1940$ MeV
and a width of around $200$ MeV.

In Fig.~\ref{Fig:angdis} (a), (b), and (c), we show the
$\Sigma^+(1385)$, neutron and $K^+$ angular distributions in the
CMS, respectively. The anisotropy of the experimental distributions
can be well reproduced thanks to the contributions from the
$\Delta^{*}(1940)$ resonance. The results obtained in the helicity
frame with respect to the angle, $\Theta^{a-b}_{c-d}$, which
represents the angel between particles ``$a$" and ``$b$" in the
``$c$" and ``$d$" reference frame (see more details in
Ref.~\cite{Agakishiev:2011qw}), are shown in Fig.~\ref{Fig:angdis}
(d), (e), and (f), while Fig.~\ref{Fig:angdis} (g), (h), and (i)
depict the distributions of the Gottfried-Jackson angles.

\begin{figure*}[htbp]
\begin{center}
\includegraphics[scale=0.6]{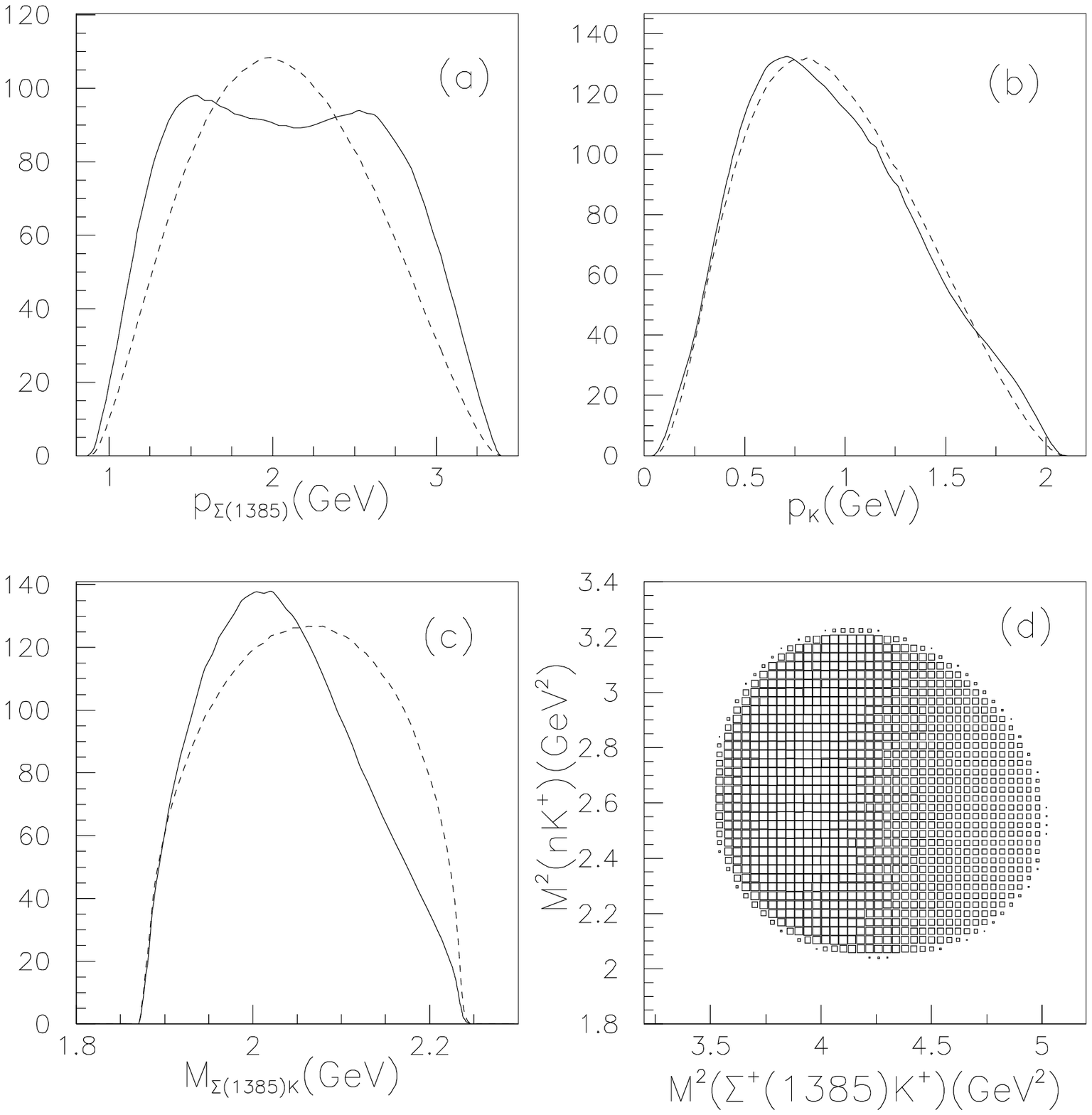}
\caption{Momentum distribution (arbitrary units), invariant mass
spectrum (arbitrary units), and Dalitz Plot for the $pp \to n K^+
\Sigma^+(1385)$ reaction at beam energy ${\rm p_{lab}} = 4.34$ GeV
comparing with the phase space distribution.The dashed lines are
pure phase space distributions, while the solid lines are full
results from our model.} \label{Fig:imdp}
\end{center}
\end{figure*}

Furthermore, the corresponding momentum distribution~\footnote{It is
noteworthy that our results are calculated in the reaction
laboratory frame, in which the target proton is at rest.} of the
$\Sigma^+(1385)$ and $K^+$ meson, the $K\Sigma(1385)$ invariant mass
spectrum, and also the Dalitz Plot for the $pp \to n K^+
\Sigma^+(1385)$ reaction at beam momentum ${\rm p_{lab}} = 4.34$ GeV
(corresponding to kinetic beam energy ${\rm T_p} = 3.5$
GeV\footnote{${\rm p_{lab}}$ = $\sqrt{{\rm E_{lab}}^2 - m^2_p}$ =
$\sqrt{({\rm T_p}+m_p)^2-m^2_p}$.}), which is accessible for HADES
Collaboration~\cite{Agakishiev:2011qw}, are calculated and shown in
Fig.~\ref{Fig:imdp}(a), Fig.~\ref{Fig:imdp}(b),
Fig.~\ref{Fig:imdp}(c), and Fig.~\ref{Fig:imdp}(d), respectively.
The dashed lines are pure phase space distributions, while, the
solid lines are full results from our model. From
Fig.~\ref{Fig:imdp}(c), we can see that at ${\rm p_{lab}} = 4.34$
GeV, our model results on the the momentum distribution of the
$\Sigma^+(1385)$ are much different with the phase space. On the
other hand, there is a clear bump in the $K\Sigma(1385)$ invariant
mass distribution, which is produced by including the contribution
from $\Delta^*(1940)$ resonance.

The momentum distribution, invariant mass spectra and the Dalitz
plots in Fig.~\ref{Fig:imdp} show direct information about the $pp
\to n K^+\Sigma^+(1385)$ reaction mechanism and may be tested by the
future experiments.

In summary, owing to the important role played by the resonant
contribution in the $pp \to n K^+ \Sigma^+(1385)$ reaction, our
model can describe the experimental data of the angle distributions
well, which indicate that recent HADES data support the existence of
this $\Delta^*(1940)$ resonance, and more accurate data for this
reaction can be used to improve our knowledge on the
$\Delta^*(1940)$ properties, which are at present poorly known. Our
present calculation offers some important clues for the mechanisms
of the $\pi^+ p \to K^+ \Sigma^+(1385)$ and $pp \to n K^+
\Sigma^+(1385)$ reactions and makes a first effort to study the role
of the $\Delta^*(1940)$ resonance in relevant reactions.

\section{Summary}

In this paper, the $\Sigma^+(1385)$ hadronic production in
proton-proton and $\pi^+ p$ collisions are studied within the
combination of the effective Lagrangian approach and the isobar
model. For $\pi^+ p \to K^+ \Sigma^+(1385)$ reaction, in addition to
the ``background" contributions from $t$-channel $K^{*0}(892)$
exchange, $u$-channel $\Sigma^0(1193)$ and $\Lambda(1115)$ exchange,
we also considered the contribution from the $\Delta^*(1940)$
resonance in the $s$-channel, which has significant coupling to
$K\Sigma(1385)$ channel. We show that the inclusion of the
$\Delta^*(1940)$ resonance leads to a fairly good description of the
low energy experimental total cross section data of $\pi^+ p \to K^+
\Sigma^+(1385)$ reaction. The $s$-channel $\Delta^*(1940)$ resonance
gives the dominant contribution, while the $t$-channel and
$u$-channel diagrams give the minor contributions.

From $\chi^2$-fit to the available experimental data for the $\pi^+
p \to K^+ \Sigma(1385)$ reaction, we get the mass and total decay
width of $\Delta^*(1940)$, which are $M_{\Delta^*} = 1940 \pm 24$
MeV and $\Gamma_{\Delta^*} = 172 \pm 94$ MeV, respectively. With the
value $0.35 \pm 0.11$ for the $\Delta^*(1940) N \pi$ coupling
constant $g_{\pi N \Delta^*}$, which is obtained with the branching
ration Br($\Delta^*(1940)\to N \pi) = (5 \pm 2) \%$, we determine
the strong couplings $g_{1,2}$ for the $\Delta^*(1940)K\Sigma(1385)$
vertex as $g_1 = -1.04 \pm 0.38$ and $g_2 = 5.24 \pm 2.30$. With
these above values, we have calculated the partial decay width of
$\Delta^*(1940) \to\Sigma(1385) K $, and we obtain $\Gamma_{\Delta^*
\to \Sigma^* K} = 17.9 \pm 12.9$ MeV and Br($\Delta^* \to \Sigma^*
K$) = ($10.4 \pm 4.9$)\%. It is shown that the $\Delta^*(1940)$
resonance would have a large decay width into $\Sigma(1385) K$,
which will be compatible with the findings of the QM approach of
Ref.~\cite{simonprd58}.

Based on the study of $\pi^+ p \to n K^+\Sigma^+(1385)$ reaction, we
study the $pp \to n K^+ \Sigma^+(1385)$ reaction with the assumption
that the production mechanism is due to the $\pi^+$-meson exchanges.
We give our predictions about total cross sections for the $pp \to n
K^+ \Sigma^+(1385)$ reaction. We find that our theoretical results
with the dominant contributions from the $\Delta^*(1940)$ resonance
can describe fairly well the experimental data both on total cross
sections and differential cross sections. Thus, the HADES data
support the important role played by the $\Delta^{*}(1940)$
resonance with a mass in the region of $1940$ MeV and a width of
around $200$ MeV. Furthermore, we also demonstrate that the
invariant mass distribution and the Dalitz Plot provide direct
information of the $pp \to n K^+ \Sigma^+(1385)$ reaction mechanisms
and may be tested by the future experiments.

Finally, we would like to stress that the $pp \to n K^+
\Sigma^+(1385)$ reaction is a new excellent source for studying
$\Delta^{*}$ resonances. And due to the important role played by the
$\Delta^*(1940)$ resonance in the $\pi^+ p \to K^+ \Sigma^+(1385)$
and $pp \to n K^+ \Sigma^+(1385)$ reactions, accurate data for these
reactions can be used to improve our knowledge on the
$\Delta^*(1940)$ properties, which are at present poorly known.

\section*{Acknowledgments}

We would like to thank Juan Nieves, Bo-Chao Liu and Xu Cao for
useful discussions. This work is partly supported by DGI and FEDER
funds, under contract No. FIS2011-28853-C02-01 and
FIS2011-28853-C02-02, the Spanish Ingenio-Consolider 2010 Program
CPAN (CSD2007-00042), Generalitat Valenciana under contract
PROMETEO/2009/0090 and by the National Natural Science Foundation of
China under grants 11105126, 11035006, 11121092, 11261130311 (CRC110
by DFG and NSFC), the Chinese Academy of Sciences under Project No.
KJCX2-EW-N01 and the Ministry of Science and Technology of China
(2009CB825200). We acknowledge the support of the European
Community-Research Infrastructure Integrating Activity "Study of
Strongly Interacting Matter" (HadronPhysics3, Grant Agreement No.
283286) under the Seventh Framework Programme of EU.


\begin{thebibliography}{99}
%
\bibitem{klempt} E. Klempt and J. M. Richard, Rev. Mod. Phys. \textbf{82}, 1095 (2010).
%
\bibitem{pdg2012} J. Beringer et al., [Particel Data Group], Phys. Rev. D \textbf{86}, 010001 (2012).
%
\bibitem{capstick2000}S. Capstick and W. Robert, Prog. Part. Nucl. Phys. \textbf{45}, S241 (2000), and references therein.

\bibitem{Gamermann:2011mq}
  D.~Gamermann, C.~Garcia-Recio, J.~Nieves and L.~L.~Salcedo,
  Phys. Rev. D {\bf 84}, 056017 (2011).

\bibitem{Sarkar:2009kx}
  S.~Sarkar, B.~-X.~Sun, E.~Oset and M.~J.~Vicente Vacas,
  Eur.\ Phys.\ J.\ A {\bf 44}, 431 (2010).

\bibitem{Oset:2009vf}
  E.~Oset and A.~Ramos,
  Eur.\ Phys.\ J.\ A {\bf 44}, 445 (2010).

\bibitem{Sun:2011fr}
  B.~-X.~Sun, H.~-X.~Chen and E.~Oset,
  Eur.\ Phys.\ J.\ A {\bf 47}, 127 (2011).
\bibitem{Xie:2007vs}
  J.~-J.~Xie and B.~-S.~Zou,
  Phys.\ Lett.\ B {\bf 649}, 405 (2007). 
\bibitem{Doring:2010ap}
  M.~Doring, C.~Hanhart, F.~Huang, S.~Krewald, U.~-G.~Meissner and D.~Ronchen,
  Nucl.\ Phys.\ A {\bf 851}, 58 (2011). 

\bibitem{Zhang:2004xt}
  A.~Zhang, Y.~R.~Liu, P.~Z.~Huang, W.~Z.~Deng, X.~L.~Chen and S.~-L.~Zhu,
  High Energy Phys.\ Nucl.\ Phys.\  {\bf 29}, 250 (2005). 
\bibitem{Wu:2009tu}
  J.~-J.~Wu, S.~Dulat and B.~S.~Zou,
  Phys.\ Rev.\ D {\bf 80}, 017503 (2009). 
\bibitem{Wu:2009nw}
  J.~-J.~Wu, S.~Dulat and B.~S.~Zou,
  Phys.\ Rev.\ C {\bf 81}, 045210 (2010). 

\bibitem{Klein:1970ri}
  S.~Klein, W.~Chinowsky, R.~R.~Kinsey, M.~Mandelkern, J.~Schultz and T.~H.~Tan,
  Phys.\ Rev.\ D {\bf 1}, 3019 (1970).  
\bibitem{Agakishiev:2011qw}
  G.~Agakishiev {\it et al.}  [HADES Collaboration],
  Phys.\ Rev.\ C {\bf 85}, 035203 (2012). 
\bibitem{Ferrari:1969fr}
  E.~Ferrari,
  Phys.\ Rev.\  {\bf 175}, 2003 (1968).  
\bibitem{Chinowsky:1968rn}
  W.~Chinowsky, R.~R.~Kinsey, S.~L.~Klein, M.~Mandelkern, J.~Schultz, F.~Martin, M.~L.~Perl and T.~H.~Tan,
  Phys.\ Rev.\  {\bf 165}, 1466 (1968).  

\bibitem{Capstick:1992uc}
  S.~Capstick,
  Phys.\ Rev.\  D {\bf 46}, 2864 (1992).

\bibitem{simonprd58} S.~Capstick, and W. Roberts, Phys.\ Rev.\ D {\bf 58}, 074011 (1998).
\bibitem{Oh:2007jd}
  Y.~Oh, C.~M.~Ko and K.~Nakayama,
  Phys.\ Rev.\ C {\bf 77}, 045204 (2008). 
\bibitem{Gao:2010hy}
  P.~Gao, J.~-J.~Wu and B.~S.~Zou,
  Phys.\ Rev.\ C {\bf 81}, 055203 (2010). 
\bibitem{Chen:2013vxa}
  Y.~-H.~Chen and B.~-S.~Zou,
  Phys.\ Rev.\ C {\bf 88}, 024304 (2013). 
\bibitem{He:2013ksa}
  J.~He,
  arXiv:1311.0571 [nucl-th].  
\bibitem{Xie:2011uw}
  J.~-J.~Xie, A.~Martinez Torres, E.~Oset and P.~Gonzalez,
  Phys.\ Rev.\ C {\bf 83}, 055204 (2011). 
\bibitem{Kamano:2009im}
  H.~Kamano, B.~Julia-Diaz, T.~-S.~H.~Lee, A.~Matsuyama and T.~Sato,
  Phys.\ Rev.\ C {\bf 80}, 065203 (2009).
\bibitem{Suzuki:2009nj}
  N.~Suzuki, B.~Julia-Diaz, H.~Kamano, T.~-S.~H.~Lee, A.~Matsuyama and T.~Sato,
  Phys.\ Rev.\ Lett.\  {\bf 104}, 042302 (2010).
\bibitem{Kamano:2008gr}
  H.~Kamano, B.~Julia-Diaz, T.~-S.~H.~Lee, A.~Matsuyama and T.~Sato,
  Phys.\ Rev.\ C {\bf 79}, 025206 (2009).

\bibitem{Kamano:2011ih}
  H.~Kamano, S.~X.~Nakamura, T.~S.~H.~Lee and T.~Sato,
  Phys.\ Rev.\ D {\bf 84}, 114019 (2011).

\bibitem{MartinezTorres:2009cw}
  A.~Martinez Torres, K.~P.~Khemchandani, U.~-G.~Meissner and E.~Oset,
  Eur.\ Phys.\ J.\ A {\bf 41}, 361 (2009).
\bibitem{alberto}
  A.~Martinez Torres, K.~P.~Khemchandani and E.~Oset,
  Phys.\ Rev.\ C {\bf 79}, 065207 (2009)
\bibitem{Zou:2002yy}
  B.~S.~Zou and F.~Hussain,
   Phys.\ Rev.\ C {\bf 67}, 015204 (2003). 

\bibitem{Kim:2012pz}
  S.~-H.~Kim, S.~-I.~Nam, A.~Hosaka and H.~-C.~Kim,
  Phys.\ Rev.\ D {\bf 88}, 054012 (2013). 
\bibitem{Xie:2013wfa}
  J.~-J.~Xie, B.~-C.~Liu and C.~-S.~An,
  Phys.\ Rev.\ C {\bf 88}, 015203 (2013). 
\bibitem{Xie:2013db}
  J.~-J.~Xie and B.~-C.~Liu,
  Phys.\ Rev.\ C {\bf 87}, 045210 (2013). 

\bibitem{Hanson:1972zza}
  P.~Hanson, G.~E.~Kalmus and J.~Louie,
  Phys.\ Rev.\ D {\bf 4}, 1296 (1971); see also http://hepdata.cedar.ac.uk/view/ins74874.

\bibitem{dagan} S. Dagan, Z. Ming Ma, J. W. Chapman, L. R. Fortney, and E. C. Fowler, Phys.\ Rev.\ {\bf 161}, 1384 (1967).

\bibitem{Butler:1973gq}
  W.~R.~Butler, D.~G.~Coyne, G.~Goldhaber, J.~Macnaughton and G.~H.~Trilling,
  Phys.\ Rev.\ D {\bf 7}, 3177 (1973).  
\bibitem{Stodolsky:1963kt}
  L.~Stodolsky and J.~J.~Sakurai,
  Phys.\ Rev.\ Lett.\  {\bf 11}, 90 (1963).
\bibitem{Stodolsky:1964zz}
  L.~Stodolsky,
  Phys.\ Rev.\  {\bf 134}, B1099 (1964).

\bibitem{cutkosky}
  R.~E.~Cutkosky, C.~P.~Forsyth, J.~B.~Babcock, R.~L.~Kelly and R.~E.~Hendrick,
  COO-3066-157. {\it Baryon} 80 Proceedings of the 4th International Conference on Baryon Resonances,
  Toronto, Canada, July 14-16, 1980, ed. N. Isgur, University of Toronto, 1981, p.19.
\bibitem{AbdelBary:2010pc}
  M.~Abdel-Bary {\it et al.}  [COSY-TOF Collaboration],
  Eur.\ Phys.\ J.\ A {\bf 46}, 27 (2010);  [Erratum-ibid.\ A {\bf 46}, 435 (2010)]. 

\end{thebibliography}
\end{document}